\documentclass[12pt]{article}
\usepackage[utf8]{inputenc}
\usepackage{graphicx}
\usepackage{animate}
\usepackage{amsmath,bm,amsfonts,amssymb,enumerate,epsfig,bbm,calc,color,ifthen,capt-of,multimedia,bm,multirow}
\usepackage{blindtext}
\usepackage{booktabs}
\usepackage{MnSymbol}
\usepackage{url}
\usepackage{natbib}
\usepackage{changepage}
\usepackage{fancyhdr}
\usepackage[flushleft]{threeparttable}
\newtheorem{theorem}{Theorem}
\usepackage{setspace}
\title{Identification and Estimation of Causal Effects with Confounders Missing Not at Random}
\author{Jian Sun and Bo Fu \thanks{Corresponding author: Bo Fu, School of Data Science,  Fudan University, Shanghai 200433, China. Email: fu@fudan.edu.cn.}\\
School of Data Science, Fudan University
 }
\date{}

\begin{document}

\maketitle

\begin{abstract}
\fontsize{12}{14pt plus.8pt minus .6pt}\selectfont

Making causal inferences from observational studies can be challenging when confounders are missing not at random. In such cases, identifying causal effects is often not guaranteed. Motivated by a real example, we consider a treatment-independent missingness assumption under which we establish the identification of causal effects when confounders are missing not at random. We propose a weighted estimating equation (WEE) approach for estimating model parameters and introduce three estimators for the average causal effect, based on regression, propensity score weighting, and doubly robust estimation. We evaluate the performance of these estimators through simulations, and provide a real data analysis to illustrate our proposed method.
\end{abstract}

\section{Introduction}
Observational studies have become essential for evaluating the causal effects of treatments or exposures on outcomes when  randomised experiments are infeasible. Estimating causal effects from observational data is a challenging task as it requires adequate control for potential confounding of the treatment-outcome relationship. Standard methods such as propensity score weighting \citep{Rosenbaum_1987}, subclassification \citep{Rosenbaum_Rubin_1984}, and matching \citep{Rosenbaum_Rubin_1983} have been proposed to adjust for confounding from baseline characteristics between treated and untreated individuals when all the confounders are fully observed.

However, confounders are often subject to missingness in practice. Different approaches have been proposed to handle unmeasured confounding for causal inference, such as instrumental variable \citep{Angrist_Imbens_Rubin_1996}, propensity score calibration \citep{Stumer_et_al_2005}, and sensitivity analysis \citep{VanderWeeele_Arah_2011}.  Nonetheless, missing data problem in partially observed confounders has received much less attention in the research literature. According to Robin's taxonomy \citep{Little_Rubin_2014}, missing values in confounders may occur under different missing data mechanisms including missing completely at random (MCAR), missing at random (MAR), and missing not at random (MNAR). When the probability of a confounder being missing does not depend on any observed or unobserved information, the confounder is MCAR. Using complete case analysis will result in an unbiased estimator of the average causal effect in such cases \citep{Imai_VanDyk_2004}. When the probability of a confounder being missing depends only on observed data values but not on missing information given observed information, the confounder is MAR. Various methods such as multiple imputation \citep{Rubin_1987, Qu_Lipkovich_2009, Crowe_et_al_2010, Mitra_Reiter_2011, Seaman_White_2014, leyrat_et_al_2019, Shan_Thomas_Gutman_2021}, fractional imputation \citep{Corder_Yang_2019},  inverse missing probability weighting \citep{Moodie_et_al_2008, Leyrat_et_al_2021}, Bayesian nonparametric generative models \citep{Roy_et_al_2017}, and doubly robust methods \citep{Williamson_Forbe_Wolfe_2012, Bagmar_Shen_2022} have been proposed to handle MAR confounders in causal effect estimation. 

When confounders are MNAR, the missing-data mechanism is dependent on unobserved information. In some special MNAR cases, non-parametric identification of causal effects is possible. \cite{Mohan_Pearl_2021} illustrated a particular situation where causal effects can be consistently estimated when the probability of a confounder being missing depends on other partially observed confounders. Nonetheless, in general, causal effects are often non-identifiable when the probability of a confounder being missing depends on unobserved values of the confounder itself \citep{Frangakis_et_al_2007, Egleston_Scharfstein_MacKenzie_2009}. Methods for handling missing covariates or confounders under the MNAR scenario have been recently proposed in the literature. For example,  \cite{Ding_Geng_2014} discussed the identifiability of causal effects in randomised experiments with missing covariates under different missing data mechanisms including MNAR for discrete covariates and outcomes. \cite{Yang_Wang_Ding_2019} considered the problem of handling MNAR confounders in causal inference from observational studies and showed that the causal effects are identifiable under a particular setting in which the missing data mechanism is independent of the outcome given the treatment and possibly missing confounders. They proposed a nonparametric two-stage least squares estimator as well as parametric likelihood-based methods for causal effect estimation. \cite{Sun_Liu_2021} proposed semiparametric estimators for the average casual effect with nonignorable missing confounders under the same outcome-independent missingness assumption as in \cite{Yang_Wang_Ding_2019}.  \cite{Yang_Lorch_Small_2014} considered the nonignorable missing covariate problem when using instrumental variable methods to control for unmeasured confounding in observational studies and suggested a maximum likelihood method implemented by EM algorithm to obtain an unbiased estimator of the causal effects. All the aforementioned methods for handling MNAR covariates or confounders required some statistically untestable assumptions to establish identifiability of causal effects and to make valid inference. Alternatively,  \cite{Lu_Ashmead_2018} proposed a sensitivity analysis approach for investigating the impact of an MNAR confounder on causal effect estimation, avoiding trouble with identification of causal effects under different missing data mechanism assumptions.

This paper was motivated by an epidemiological study that examined the potential bias resulting from missing values of a confounding variable when estimating the causal effect of marital status on the outcome of depression \citep{Knol_et_al_2010}. Baseline age, gender, and income were considered as potential confounders that affect both the exposure and the outcome. The study created missing values in the confounding variable income according to two missingness mechanisms: MCAR and MAR. It demonstrated that commonly used methods to handle missing confounder data, such as complete-case analysis and the missing indicator method, may introduce unpredictable bias into the effect estimate. In contrast, multiple imputation give unbiased effect estimates when missing values are MAR. However,  \cite{Davern_et_al_2005} suggested that the probability of self-reported income data being missing was usually related to the value of income itself, implying that the underlying missing data mechanism is more likely to be MNAR in this example. Since the income data were collected via questionnaires, their missingness may also be related to the mental health outcome. Therefore, the approach proposed by \cite{Yang_Wang_Ding_2019} for handling confounder MNAR based on an outcome-independent missingness assumption is not applicable in this case. On the other hand, it is reasonable to assume that the missingness mechanism for the self-reported income data is independent of marital status, given the outcome and all possible confounders including income. This motivates us to propose a novel framework for identifying and estimating causal effects with MNAR confounders based on the treatment-independent missingness assumption. A similar assumption was discussed in \cite{Ding_Geng_2014} for establishing identifiability of causal effects in randomised experiments with MNAR covariates. Our work is also linked to the shadow variable assumption introduced by \cite{Miao_Tchetgen_Tchetgen_2018}, which employs a continuous shadow variable to identify parametric models in cases where covariates are MNAR. However, in our study, the treatment variable, which corresponds to the shadow variable in \cite{Miao_Tchetgen_Tchetgen_2018}'s work, is binary rather than continuous. As a result, it contains less information, making identification more challenging.  Furthermore, our focus is on causal inference, and the partially observed confounder is a cause of the treatment variable in our model, whereas the partially observed covariate is a child node of the shadow variable in the directed acyclic graph of \cite{Miao_Tchetgen_Tchetgen_2018}. This leads to different modelling of the conditional distributions for the treatment variable or the partially observed variable. 

The rest of this paper is organized as follows. Section 2 presents a general strategy for identifying causal effects. In Section 3 we propose a weighted estimating equation approach for estimating model parameters and three estimators for the average casual effect based on outcome regression modelling, inverse probability weighting and double robust idea. Section 4 presents simulation studies to evaluate our proposed method and compare its performance with existing methods. The proposed estimators are illustrated by a real data analysis in Section 5. 

\section{Identification}

\subsection{Notation and Assumptions}

We consider a binary treatment variable denoted by $A$, where $A=0$ and $A=1$ correspond to the control and treatment groups, respectively. Let $\bm{C} = (C_{1},C_{2},\ldots, C_{m})^{\mathrm{T}}$ be a vector of $m$-dimensional confounders and $Y$ denote the outcome of interest. To avoid ambiguity, we use capital letters to represent random variables and lowercase letters to indicate specific realizations of random variables. For each subject, there exists a pair of potential outcomes $(Y(1), Y(0))$, where $Y(a)$ is the potential outcome for the subject that would be observed if he or she were assigned to treatment $a$. The average treatment effect is represented by $\tau = E[Y(1)-Y(0)]$. To identify $\tau$, we usually require the causal consistency assumption 
\begin{equation}
    Y=AY(1)+(1-A)Y(0), \label{2.1} \tag{2.1}
\end{equation}
the overlap assumption
\begin{equation}
    0<P(A=1\mid \bm{C})<1, \label{2.2} \tag{2.2}
\end{equation}
and the unconfoundedness assumption
\begin{equation}
    \{Y(1),Y(0)\} \upmodels A\mid \bm{C}. \label{2.3} \tag{2.3}
\end{equation}
In the absence of missing data, standard propensity score methods (such as matching or weighting) can be used to estimate the causal effects \citep{Hernan_Robins_2020}.

For simplicity, we focus on the case where there is only one partially observed confounder. Specifically, let $C_1$  be the partially observed confounder and $R_C$ denote the missing indicator of $C_1$. The value of $R_C$ is equal to $1$ when $C_1$ is observed and $0$ when it is missing. Thus, we observe $(A,Y,R_C,C_2,\ldots,C_m)^{\mathrm{T}}$ for all subjects, and $C_1$ is only available for those with $R_C=1$.  We make the following assumption to avoid degeneracy of the missing data mechanism:
\begin{equation}
    P(R_C=1\mid A,\bm{C},Y) > 0.  \label{2.4} \tag{2.4}
\end{equation} 
To establish identification, we consider an additional treatment-independent missingness assumption for the missing data mechanism, that is,  
\begin{equation}
R_C \upmodels A \mid (\bm{C},Y),  \label{2.5} \tag{2.5}
\end{equation}
which states that given the outcome $Y$ and confounders $\bm{C}$, the missing indicator $R_C$ is conditional independent of the treatment.

\begin{figure}[htbp]
\centering
\includegraphics[height = 3.5cm]{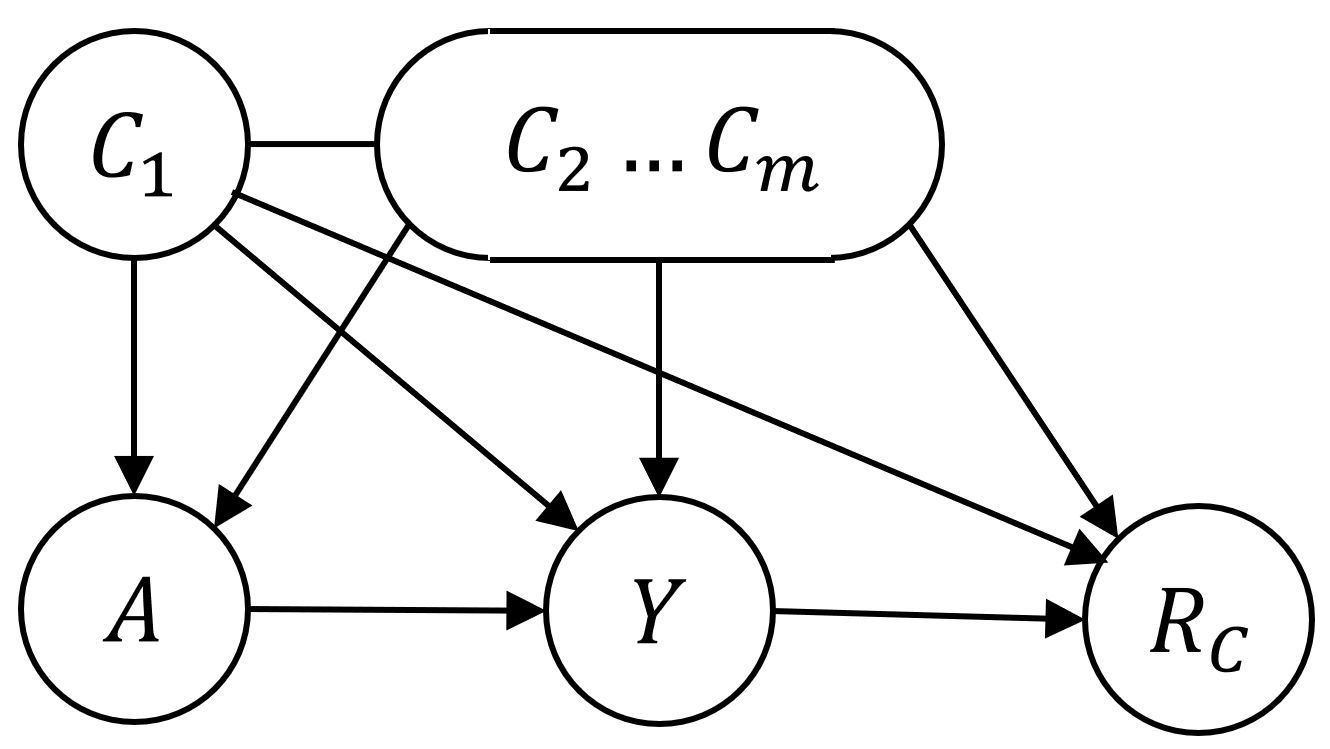}
\linespread{1}
\caption{A DAG illustrating the treatment-independent missingness assumption}
\label{fig1}
\end{figure}

Assumptions (2.1)-(2.5) make it possible to identify the causal effects with a confounder MNAR. The causal relationships between these variables are depicted in Figure \ref{fig1}. 

Notably, the treatment variable $A$ can be seen as a shadow variable, as discussed by \cite{Miao_Tchetgen_Tchetgen_2016}, which has been employed in the literature on outcome MNAR \citep{D’Haultfoeuille_2010, Wang_et_al_2014, Zhao_Shao_2015, Miao_Tchetgen_Tchetgen_2016, Zhao_Ma_2022, Li_et_al_2023}. While previous studies have utilized a continuous shadow variable to address identification challenges in the presence of nonignorable missing covariate data \citep{Miao_Tchetgen_Tchetgen_2018}, the treatment variable $A$ in Figure 1 is binary, which inherently carries less information compared to a continuous shadow variable. It is noteworthy that we neither assume independence among the confounders $(C_{1},\ldots, C_{m})$ nor impose constraints on the marginal and conditional distributions of the partially observed confounder. Instead, additional assumptions regarding the outcome model, the treatment propensity score model, and the missing data model are required to ensure the identification of causal effects. Further elaboration on these assumptions will be provided in Section 2.3 of the paper. Furthermore, it is important to highlight that the identifiability of certain parameters may differ when employing a binary shadow variable as opposed to a continuous one, even within the same data-generation mechanism. To illustrate this point, we present an example as follows. 

\textbf{Example 1.}  Let $C\sim N(\eta_1 A,\lambda)$ and $Y~\sim N(A,1)$. Consider a logistic missing probability model: $\operatorname{logit}P(R_C = 1\mid c) = \alpha_0+\alpha_1 c^2$. According to the Theorem 2 in \cite{Miao_Tchetgen_Tchetgen_2018}, if the treatment variable $A$ were continuous, all parameters would be identifiable. However, if $A\in\{-1,1\}$ with $P(A=1) = 0.5$, for two different sets of parameters: $ (\eta_1,\lambda,\alpha_0,\alpha_1) = (1,1,-(1+\ln 2)/2,1/4)$ and $ (\eta_1',\lambda',\alpha_0',\alpha_1') = (2,2,(1+\ln 2)/2,-1/4)$, we can demonstrate that they lead to the same observed data distribution. Consequently, these parameters are not identifiable. The details for this example are provided in Web Appendix A.

\subsection{A General Identification Strategy}

Consider a model pr$(a, \bm{c}, y, r_c; \bm{\theta})$ indexed by $\bm{\theta}$, where $\bm{c} = (c_1,c_2,\ldots,c_m)^\mathrm{T}$ is the vector of all confounders and $\bm{\theta} = (\theta_1, \theta_2, \ldots, \theta_p)^\mathrm{T}$ is the vector of all parameters. A parameter  $\theta_i$ is considered identifiable if and only if the mapping from the parameter space $\Theta_i = \{\theta_i\}$ to the space of the observed data distribution $\{ \operatorname{pr}(a,\bm{c},y,R_C=1;\bm{\theta}), \operatorname{pr}(a,c_2,\ldots,c_m,y,R_C=0;\bm{\theta}); \bm{\theta}\in \bm{\Theta} \}$ is injective. Note that there is no constraint on other parameters except for $\theta_i$. Thus, it is possible that for two parameters $\theta_{i1}, \theta_{i2} \in \bm{\theta}$, $\theta_{i1}$ is identifiable while $\theta_{i2}$ is not.

\

 \textbf{Proposition 1.}\quad Consider two parameter sets $\bm{\theta} = (\theta_1,\theta_2,\ldots,\theta_p)^\mathrm{T}$, $ \bm{\theta}^{\prime} =(\theta_1^{\prime}, \theta_2^{\prime},\ldots,\theta_p^{\prime})^\mathrm{T}$ with 
$\theta_i \neq \theta_i^{\prime}$ $ (i \in \{1,2,\ldots,p\})$. If for any function $h(\bm{c}, y)$, $\exists \mathcal{F}$, $P(\mathcal{F})\neq 0$ and for all $(a,\bm{c},y) \in \mathcal{F}$, $\frac{\operatorname{pr}\left(a,y \mid \bm{c} ; \bm{\theta} \right)}{ \operatorname{pr}\left(a,y \mid \bm{c}; \bm{\theta^{\prime}}\right)} \neq h(\bm{c}, y)$, then the parameter $\theta_i$ is identifiable.

\textbf{Proof}. Under the treatment-independent missingness assumption (2.5), we have
\begin{align*}
     \frac{\operatorname{pr}(R_C=1 \mid a,\bm{c},y;\bm{\theta^{\prime}})}{ \operatorname{pr}(R_C=1 \mid a,\bm{c},y;\bm{\theta})} &= \frac{ \operatorname{pr}(R_C=1 \mid \bm{c},y;\bm{\theta^{\prime}})}{ \operatorname{pr}(R_C=1 \mid \bm{c},y;\bm{\theta})},
\end{align*}
implying that $\frac{\operatorname{pr}(\bm{c};\bm{\theta^{\prime}})}{\operatorname{pr}(\bm{c};\bm{\theta})} \frac{\operatorname{pr}(R_C=1 \mid a,\bm{c},y;\bm{\theta^{\prime}})}{ \operatorname{pr}(R_C=1 \mid a,\bm{c},y;\bm{\theta})}$ is a function of $\bm{c}, y$. It follows that if for any function $h(\bm{c}, y)$,  $\exists \mathcal{F}$, $P(\mathcal{F})\neq 0$ and for all $(a,\bm{c},y) \in \mathcal{F}$, $\frac{\operatorname{pr}\left(a,y \mid \bm{c} ; \bm{\theta} \right)}{ \operatorname{pr}\left(a,y \mid \bm{c}; \bm{\theta^{\prime}}\right)} \neq h(\bm{c}, y)$, then the following inequality holds with a positive probability:
\begin{align*}
    \frac{\operatorname{pr}(a,y \mid \bm{c};\bm{\theta})}{\operatorname{pr}(a,y \mid \bm{c};\bm{\theta^{\prime}})} &\neq \frac{\operatorname{pr}(\bm{c};\bm{\theta^{\prime}})\cdot \operatorname{pr}(R_C=1 \mid a,\bm{c},y;\bm{\theta^{\prime}})}{\operatorname{pr}(\bm{c};\bm{\theta})\cdot \operatorname{pr}(R_C=1 \mid a,\bm{c},y;\bm{\theta})},
\end{align*}    
that is, 
\begin{align*}
\operatorname{pr}(a,\bm{c},y,R_C=1;\bm{\theta}) &\neq \operatorname{pr}(a,\bm{c},y,R_C=1;\bm{\theta^{\prime}}).
\end{align*}    
Therefore, the observed data distributions can not be identical when $\theta_i \neq \theta_i^{\prime}$, which means that $\theta_i$ is identifiable.

Proposition 1 indicates that if the ratio of two different $\operatorname{pr}(a,y \mid \bm{c}; \bm{\theta})$ models indexed by different parameter values $\theta_i$ and $\theta_i^{\prime}$ varies with $a$, then $\theta_i$ is identifiable. Since the conditional distribution $\operatorname{pr}(a,y \mid \bm{c})$ can be determined by the treatment propensity score model $\operatorname{pr}(a\mid \bm{c})$ and the outcome model $\operatorname{pr}(y\mid a,\bm{c})$, Proposition 1 implies that under the treatment-independent missingness assumption, it may be possible to identify certain parameters without requiring additional assumptions regarding the missing probability model. See for example the results in Theorem 3(c) in Section 2.3.

\

 \textbf{Proposition 2.}\quad If $\theta_i = \theta_i^{\prime}$ is a necessary condition for the following equation to hold for all $y, \bm{c}$:
\begin{center}
$\frac{\operatorname{pr}\left(y \mid A=1, \bm{c}  ; \bm{\theta} \right)\operatorname{pr}\left(A=1\mid \bm{c}  ; \bm{\theta} \right)}{\operatorname{pr}\left(y \mid A=0, \bm{c}  ; \bm{\theta} \right)\operatorname{pr}\left(A=0\mid \bm{c}  ; \bm{\theta} \right)} =  \frac{\operatorname{pr}\left(y \mid A=1, \bm{c}  ; \bm{\theta}^{\prime} \right)\operatorname{pr}\left(A=1\mid \bm{c}  ; \bm{\theta}^{\prime} \right)}{\operatorname{pr}\left(y \mid A=0, \bm{c}  ; \bm{\theta}^{\prime} \right)\operatorname{pr}\left(A=0\mid \bm{c}  ; \bm{\theta}^{\prime} \right)}$,    
\end{center}
then $\theta_i$ is identifiable

\textbf{Proof}. According to Proposition 1, $\theta_i$ is identifiable if $\frac{\operatorname{pr}\left(a,y \mid \bm{c}  ; \bm{\theta} \right) }{ \operatorname{pr}\left(a,y \mid \bm{c}  ; \bm{\theta^{\prime}}\right)}$ varies with $a$ for any $\theta_i \neq \theta_i^{\prime}$. Because $A$ is binary, $\theta_i$ is identifiable if  $\frac{\operatorname{pr}\left(A=0,y \mid \bm{c}  ; \bm{\theta} \right)}{\operatorname{pr}\left(A=0,y \mid \bm{c}  ; \bm{\theta^{\prime}} \right)}\neq\frac{\operatorname{pr}\left(A=1,y \mid \bm{c}  ; \bm{\theta} \right)}{\operatorname{pr}\left(A=1,y \mid \bm{c}  ; \bm{\theta^{\prime}} \right)}$ with a positive probability. Proposition 2 then follows by using $\operatorname{pr}\left(a,y \mid \bm{c} \right) = \operatorname{pr}\left(y \mid a,\bm{c} \right) \cdot \operatorname{pr}\left(a \mid \bm{c} \right)$ and proof by contraposition. 

The propositions provide sufficient conditions for identifying model parameters under the treatment-independent missingness assumption, and they are useful for verifying identification of causal effects or data distributions with specified parametric models in the subsequent sections.

\subsection{Identification with Parametric Assumptions}

In this subsection, we utilize the propositions given in Section 2.2 to establish the identification for some commonly-used parametric models. To ensure the identification of the full data distribution $\operatorname{pr}(a,\bm{c},y)$, we may require additional constraints on the missing probability model:
\begin{equation}
   \operatorname{logit}\{\operatorname{pr}(R_C=1 \mid \bm{c},y)\} = M_c(\bm{c};\bm{\alpha_c})+\alpha_y y, \quad \alpha_y \neq 0, \label{2.6} \tag{2.6}
\end{equation}
where $M_c(\bm{c};\bm{\alpha_c})$ can be any specific function of $\bm{c}$ with unknown parameters $\bm{\alpha_c}$ and $\alpha_y$. Model (2.6) implies that the missing probability is a monotone function of $y$. 

We first consider the identification for Gaussian-distributed outcome models.

\begin{theorem}
For the outcome model $Y \mid a,\bm{c} \sim N(O_{a,c}(a, \bm{c}; \bm{\beta}), \phi)$, where $O_{a,c}(\cdot)$ can be any specific function of $(a, \bm{c})$ with unknown parameters $\bm{\beta}$, under assumptions (2.1)-(2.5), we have
\begin{enumerate}[(a)]
    \item if $\lim\limits_{y\to \infty}\operatorname{pr}(R_C=1|\bm{c},y)^{\frac{1}{y^2}} = 1$ and $\operatorname{pr}(R_C=1|\bm{c},y)$ is twice differentiable in $y$ for any $\bm{c}$, the conditional average treatment effect $E[Y(1)-Y(0)\mid c]$ is identifiable;
    \item if the missing probability model is (2.6) and the sign of $\alpha_y$ is known, the full data distribution is identifiable;
    \item if the missing probability model is (2.6), the full data distribution is identifiable if $\exists j \in \{1,2,\ldots,m\}, \bm{c_0}$, such that $\frac{\partial M_c(\bm{c};\bm{\alpha_c})}{\partial c_j} \mid _{\bm{c} = \bm{c_0}} \neq 0$ and the sign of $\frac{\partial M_c(\bm{c};\bm{\alpha_c})}{\partial c_j} \mid _{\bm{c} = \bm{c_0}}$ is known.
\end{enumerate}
\end{theorem}
The proof of Theorem 1 is given in the Web Appendix B. It should be noted that model (2.6) satisfies the conditions given in Theorem 1(a). The identification of the conditional average treatment effect is relatively easy, while we require more conditions in Theorem 1(b) and 1(c) for the identification of the full data distribution in order to identify the average treatment effect. In practice, we usually have prior information on the sign of $\alpha_y$ or the sign of $\frac{\partial M_c(\bm{c};\bm{\alpha_c})}{\partial c_j} \mid _{\bm{c} = \bm{c_0}}$, as domain knowledge may guide us on how the missing probability is influenced by the outcome or a confounder.

\emph{Remark 1.} \quad The commonly used linear regression for the outcome $Y \mid a,\bm{c} \sim N(\beta_0 + \beta_1 a + \bm{\beta_c}^{\mathrm{T}} \bm{c}, \phi)$ and logistic regression for the missing probability $\operatorname{logit}\{\operatorname{pr}(R_C=1 \mid \bm{c},y)\} = \alpha_0 + \bm{\alpha_c}^{\mathrm{T}} \bm{c}+ \alpha_y y$ satisfy the parametric restrictions in Theorem 1. The average treatment effect $\beta_1$ is identifiable if these models are correctly specified. 

The following example shows that the full data distribution may not be identifiable even if all the conditions required in Theorem 1(a) are satisfied, which highlights the requirement for stronger assumptions in Theorem 1(b) and 1(c).

\textbf{Example 2.} \quad Let $\operatorname{expit}(x)=\exp(x)/(1+\exp(x))$. Consider $C \sim N(\eta,1)$, $\operatorname{pr}(A=1\mid c) = \operatorname{expit}(c^2-1)$, $Y\mid a,c \sim N(\beta_0+\beta_1a|c|,\phi)$, and $\operatorname{pr}(R_C=1\mid c,y) = \operatorname{expit}(\alpha_1 c)\operatorname{expit}(\alpha_2 y)$. It is easy to show that they satisfy the conditions given in Theorem 1(a), so $\beta_1$ is identifiable. Also, we can verify that the observed data distributions are identical under the two parameter sets  $(\eta,\beta_0,\beta_1,\phi,\alpha_1,\alpha_2) = (1,0,1,1,-2,1)$ and $(\eta^{\prime},\beta_0^{\prime},\phi^{\prime},\beta_1^{\prime},\alpha_1^{\prime},\alpha_2^{\prime}) = (-1,0,1,1,2,1)$. Therefore, the full data distributions are not identifiable from the observed data.  The details for this example are provided in Web Appendix C.

We consider another scenario where the error terms in the outcome model are non-Gaussian and skewed-distributed. 
\begin{theorem}
For the model $\log(Y) = O_{a,c}(a,\bm{c}; \bm{\beta}) + \sigma W$, $\sigma \neq 0$, where  $O_{a,c}(\cdot)$ can be any specific function of $(a, \bm{c})$ with unknown parameter $\bm{\beta}$ and $W$ follows the standard extreme value distribution with probability density function $f(w) = \exp[-w-\exp(-w)]$, under assumptions (2.1)-(2.5), we have 
\begin{enumerate}[(a)]
    \item  $\sigma$ and $e^{\frac{O_{a,c}(1,\bm{c};\bm{\beta})}{\sigma}}-e^{\frac{O_{a,c}(0,\bm{c};\bm{\beta})}{\sigma}}$ are identifiable;
    \item if the missing probability model is (2.6) and $\sigma \neq -1$, the full data distribution is identifiable.
\end{enumerate}
\end{theorem}
The proof of Theorem 2 is given in Web Appendix D.

{\em Remark 2.}\quad If an exponential regression model of the form $\log(Y) = \beta_0 + \beta_1 a + \bm{\beta_c}^{\mathrm{T}} \bm{c} + W$ \citep{Lawless_2002} is correctly specified and $W\upmodels (A,\bm{C})$, Theorem 2(a) shows that the conditional average treatment effect, which equals to $e^{O_{a,c}(1,\bm{c};\bm{\beta})}-e^{O_{a,c}(0,\bm{c};\bm{\beta})}$ in this scenario, is identifiable. 

We now consider the scenario for a binary outcome. 
\begin{theorem}
For the binary outcome model $\operatorname{logit} \left\{ \operatorname{pr}(Y=1 \mid a,\bm{c})\right\} = O_{a,c}(a,\bm{c};\bm{\beta})$, $\sigma\neq 0$, where $O_{a,c}(\cdot)$ can be any specific function of $(a, \bm{c})$ and $\bm{c}$ are continuous confounding variables, under assumptions (2.1)-(2.5), we have
\begin{enumerate}[(a)]
\item the conditional causal odds ratio, defined as $\frac{\operatorname{pr}(Y=1\mid A=1,\bm{c})\operatorname{pr}(Y=0\mid A=0,\bm{c})}{\operatorname{pr}(Y=1\mid A=0,\bm{c})\operatorname{pr}(Y=0\mid A=1,\bm{c})}$, is identifiable;
\item if $\operatorname{logit} \left\{ \operatorname{pr}(Y=1 \mid a,\bm{c})\right\} = \beta_0 + \bm{\beta_c}^{\mathrm{T}}\bm{c} + a O_c(\bm{c};\bm{\beta})$, where $O_{c}(\cdot)$ can be any known function of $\bm{c}$, and $\operatorname{logit}\{\operatorname{pr}(R_C=1 \mid \bm{c},y)\} = \alpha_0+\bm{\alpha_c}^\mathrm{T}\bm{c}+\alpha_y y$, $\alpha_y \neq 0$, the full data distribution is identifiable;
\item if $\operatorname{logit} \left\{ \operatorname{pr}(Y=1 \mid a,\bm{c})\right\} = \beta_0 + \beta_1 a + \bm{\beta_c}^\mathrm{T}\bm{c}$, $\beta_1 \neq 0$, and $\operatorname{logit} \left\{ \operatorname{pr}(A=1 \mid \bm{c})\right\} = \gamma_0 + \bm{\gamma_c}^{\mathrm{T}}\bm{c}$, the outcome model and the treatment propensity score model are identifiable.
\end{enumerate}
\end{theorem}
The proof of Theorem 3 is given in Web Appendix E. \cite{Ding_Geng_2014} gave the details on how to establish identification for nonparametric causal effects when $\bm{c}$ are discrete variables. Therefore, we focus on continuous $\bm{c}$ in the binary outcome model given in Theorem 3.

\emph{Remark 3}. \quad If the commonly used logistic model $\operatorname{logit} \left\{ \operatorname{pr}(Y=1 \mid a,\bm{c})\right\} = \beta_0 + \beta_1 a + \bm{\beta_c}^\mathrm{T}\bm{c}$ is correctly specified, Theorem 3(a) shows that the causal odds ratio $\beta_1$ is identifiable. 

\section{Estimation}

In this section, we propose a two-stage approach to estimate causal effects, which involves the estimation of parameters for the missing probability model, the treatment propensity score model, and the outcome model in the first stage. In the second stage, we leverage these parametric models to develop outcome regression (OR), inverse propensity score (IPW), and doubly-robust (DR) estimators for computing the average causal effects. It should be noted that the proposed estimation methods rely on the identification of corresponding parameters.

\subsection{Weighted Estimating Equation for Parametric Models}

The inverse missing probability weighted (IPW) estimating equation is a commonly used approach for handling missing data when covariates are MAR. \cite{Miao_Tchetgen_Tchetgen_2018} and \cite{Sun_Liu_2021}  have proposed extensions of the IPW estimating equation to handle MNAR data under certain identification conditions. The IPW estimating equation is of the form $\widehat{E} \left[ \left(\frac{r_c}{\operatorname{pr}(R_C=1 \mid \bm{c},y; \hat{\bm{\alpha}})} -1 \right) G(\bm{c},y) \right]  = 0$, where $G(\bm{c},y)$ is a user-specified vector function with the same dimension as unknown parameters $\bm{\alpha}$, and $\widehat{E}$ denotes the empirical expectation. The resulting estimator is unbiased if the missing probability model is correctly specified. If $E\left[ \frac{\partial \left\{r_c/\operatorname{pr}(R_C=1\mid \bm{c},y; \hat{\bm{\alpha}})\right\} }{\partial \bm{\alpha}} G(\bm{c},y)\right]$ is nonsingular for all $\bm{\alpha}$, we can obtain consistent and asymptotically Gaussian-distributed estimators for $\bm{\alpha}$. 

However, in the presence of a MNAR confounder, this approach is not feasible, as the confounder $C_1$ is only observed when $R_C=1$. To address this issue, we replace $G(\bm{c},y)$ with $G(\bm{c_r},a,y)$ in the estimating equation:
\begin{equation}
    \widehat{E} \left[ \left(\frac{r_c}{\operatorname{pr}(R_C=1 \mid \bm{c},y; \hat{\bm{\alpha}})} -1 \right) G(\bm{c_r},a,y) \right]  = 0,  \tag{3.1} \label{3.1}
\end{equation}
where $\bm{c_r}$ denotes the fully observed confounders that are directly correlated with the partially observed confounder $C_1$, and $G(\bm{c_r},a,y)$ is a user-specified differentiable vector function of  the same dimension as $\bm{\alpha}$. For example, if $\operatorname{pr}(R_C=1 \mid \bm{c},y; \hat{\bm{\alpha}}) = \operatorname{logit} (\alpha_0 + \alpha_1 c_1 + \alpha_2 y)$, then we can set $G(\bm{c_r},a,y) = (1,a,y)^{T}$. The choice of $G(\bm{c_r},a,y)$ can affect the statistical efficiency of the estimator, and the optimal choice of $G(\bm{c_r},a,y)$ is discussed in Web Appendix F.

Similarly, we can solve the following equations to estimate the parameters in the specified treatment propensity score model and the outcome regression model:
\begin{align}
         \widehat{E} \left[ \frac{r_c}{\operatorname{pr}(R_C=1 \mid \bm{c},y; \hat{\bm{\alpha}})}   \frac{\partial \log(\operatorname{pr}(a\mid \bm{c}; \bm{\gamma}))}{\partial \bm{\gamma}} \right]  &= 0, \tag{3.2} \label{3.2}\\
        \widehat{E} \left[ \frac{r_c}{\operatorname{pr}(R_C=1 \mid \bm{c},y; \hat{\bm{\alpha}})} \frac{\partial \log(\operatorname{pr}(y\mid a,\bm{c}; \bm{\beta}))}{\partial \bm{\beta}} \right]  &= 0 . \tag{3.3} \label{3.3}
\end{align}

\begin{theorem}
 If the missing probability model $\operatorname{pr}(R_C=1 \mid \bm{c},y; \bm{\alpha})$ is correctly specified, then (\ref{3.1}) are unbiased estimating equations for $\bm{\alpha}$; if both $\operatorname{pr}(R_C=1 \mid \bm{c},y; \bm{\alpha})$ and $\operatorname{pr}(a\mid \bm{c}; \bm{\gamma})$ are correctly specified,  then (\ref{3.2}) are unbiased estimating equations for $\bm{\gamma}$; if both $\operatorname{pr}(R_C=1 \mid \bm{c},y; \bm{\alpha})$ and $\operatorname{pr}(y\mid a,\bm{c}; \bm{\beta})$ are correctly specified, then (\ref{3.3}) are unbiased estimating equations for $\bm{\beta}$; the estimators for the parameters $\bm{\alpha}$, $\bm{\beta}$ and $\bm{\gamma}$ are consistent and asymptotically Gaussian-distributed under suitable regularity conditions, that is, when $n \to \infty$,
\begin{align*}
    \sqrt{n} (\hat{\bm{\alpha}} - \bm{\alpha_{*}}) \to N(0,\bm{V_{G;\alpha}}),\\
    \sqrt{n} (\hat{\bm{\beta}} - \bm{\beta_{*}}) \to N(0,\bm{V_{G;\beta}}),\\
    \sqrt{n} (\hat{\bm{\gamma}} - \bm{\gamma_{*}}) \to N(0,\bm{V_{G;\gamma}}),
\end{align*}
where $\bm{\alpha_{*}}$, $\bm{\beta_{*}}$, and $\bm{\gamma_{*}}$ are the true values of $\bm{\alpha}$, $\bm{\beta}$, and $\bm{\gamma}$, respectively. 
\end{theorem}
The proof of Theorem 4 and the details of $\bm{V_{G;\alpha}}$, $\bm{V_{G;\beta}}$ and $\bm{V_{G;\gamma}}$ are given in Web Appendix G. The regularity conditions for the asymptotic normality in Theorem 4 can be formulated by applying the general theory of estimating equations such as Theorem 3.4 in \cite{Newey_McFadden_1994}.

\subsection{OR, IPW, and DR Estimators}

The proposed average treatment effect estimators in this section are based on the aforementioned model parameter estimators in Section 3.1. When both the outcome model and the missing probability model are correctly specified, we can utilize the estimated outcome regression functions to calculate the conditional average treatment effect. Then, the average treatment effect is estimated by averaging the conditional average treatment effect over the weighted empirical distribution of the confounding variables. The resulting estimator is defined as
\begin{align*}
    \widehat{\tau}_{OR} = \frac{1}{n} \sum_{k=1}^{n} \left\{ \frac{r_{c_k}}{M(\bm{c}_k,y_k;\hat{\bm{\alpha}})} \left[ O_{a,c}(1,\bm{c}_k; \hat{\bm{\beta}}) - O_{a,c}(0,\bm{c}_k; \hat{\bm{\beta}})\right] \right\},
\end{align*}
where $O_{a,c}(a,\bm{c}; \hat{\bm{\beta}})$ and $M(\bm{c},y;\hat{\bm{\alpha}})$ are the outcome model and the missing probability model with parameters estimated by the weighted estimating equations (WEE) approach proposed in Section 3.1, respectively. This estimator is referred to as the WEE-OR estimator. 
\begin{theorem}
If the missing probability model and the outcome model are correct, then $\widehat{\tau}_{OR}$ is a consistent estimator for the average treatment effect $\tau$ under suitable regularity conditions.
\end{theorem}
The proof of this theorem is given in Web Appendix H.

Inverse treatment propensity score weighting (IPW) is a widely used approach for estimating the average causal effect, which creates a balanced pseudo population by weighting each sample with the inverse of its conditional probability of receiving the treatment given confounders (i.e., $\operatorname{pr}(A\mid c)$). In this population, the treatment $A$ is considered to be completely randomised, enabling the difference between the average outcomes in the treatment and control groups to be an unbiased estimator of the average causal effect. The consistency of the IPW estimator does not rely on the correct specification of the outcome model. Thus, we propose the following WEE-based IPW estimators for the counterfactual outcome and the average treatment effect by employing correctly specified treatment propensity score and missing probability models, even if the outcome model is incorrect.
\begin{align*}
    \widehat{Y}(1)_{IPW} &= \frac{1}{n} \sum_{k=1}^n \left\{ \frac{r_{c_k}}{M(\bm{c}_k,y_k;\hat{\bm{\alpha}})}\frac{a_k y_k}{H(\bm{c}_k; \hat{\bm{\gamma}})}  \right\},\\
    \widehat{Y}(0)_{IPW} &= \frac{1}{n} \sum_{k=1}^n \left\{ \frac{r_{c_k}}{M(\bm{c}_k,y_k;\hat{\bm{\alpha}})}\frac{(1-a_k) y_k}{1-H(\bm{c}_k; \hat{\bm{\gamma}})} \right\},\\
    \widehat{\tau}_{IPW} &= \widehat{Y}(1)_{IPW} - \widehat{Y}(0)_{IPW},
\end{align*}
where $H(\bm{c}; \hat{\bm{\gamma}})$ is the treatment propensity score model with parameters estimated by the WEE method. We refer to this estimator as the WEE-IPW estimator.
\begin{theorem}
If the missing probability model and the treatment propensity score model are correct, then $\widehat{Y}(a)_{IPW}$ and $\widehat{\tau}_{IPW}$ are consistent estimators under suitable regularity conditions.
\end{theorem}
The proof of this theorem is provided in Web Appendix I. 

The doubly robust estimator was proposed as an augmented inverse probability weighting estimator by \cite{Robins_Rotnitzky_Zhao_1994}.
In causal inference, this method employs both the treatment propensity score model and the outcome model to produce an estimator that remains consistent if either or both of the two models is correctly specified. Additionally, the doubly robust estimator can achieve the semiparametric efficiency bound if both models are correctly specified. Using the parameters estimated by the WEE method, we can construct the doubly robust (WEE-DR) estimators as follows:
\begin{align*}
    \widehat{Y}(1)_{DR} &= \frac{1}{n} \sum_{k=1}^n \left\{ \frac{r_{c_k}}{M(\bm{c}_k,y_k;\hat{\bm{\alpha}})}  \times  \left[\frac{a_k y_k}{H(\bm{c}_k; \hat{\bm{\gamma}})} - \frac{a_k-H(\bm{c}_k; \hat{\bm{\gamma}})}{H(\bm{c}_k; \hat{\bm{\gamma}})} O_{a,c}(1,\bm{c}_k; \hat{\bm{\beta}}) \right] \right\} ,\\
    \widehat{Y}(0)_{DR}  &= \frac{1}{n} \sum_{k=1}^n \left\{ \frac{r_{c_k}}{M(\bm{c}_k,y_k;\hat{\bm{\alpha}})} \times \left[\frac{(1-a_k) y_k}{1-H(\bm{c}_k; \hat{\bm{\gamma}})} - \frac{a_k-H(\bm{c}_k; \hat{\bm{\gamma}})}{1-H(\bm{c}_k; \hat{\bm{\gamma}})} O_{a,c}(0,\bm{c}_k; \hat{\bm{\beta}}) \right]\right\},\\
    \widehat{\tau}_{DR} &= \widehat{Y}(1)_{DR} - \widehat{Y}(0)_{DR},
\end{align*}

A theorem regarding the consistency of $\widehat{Y}(a)_{DR}$ and $\widehat{\tau}_{DR}$ as estimators for $E[Y(a)]$ and $\tau$ is given below.

\begin{theorem}
  If the missing probability model is correctly specified and suitable regularity conditions are satisfied, then $\widehat{Y}(a)_{DR}$ and $\widehat{\tau}_{DR}$ are consistent estimators if one or both of the treatment propensity score model and the outcome model is correctly specified.
\end{theorem}
The proof of this theorem is given in Web Appendix J.

\section{Simulation Studies}

We conducted simulation studies to evaluate the finite-sample performance of the proposed weighted estimating equation (WEE) approach, as described in Section 3, under various scenarios. To provide a comparison, we also assessed two commonly used methods for handling missing confounder data: complete case (CC) analysis and multiple imputation (MI) \citep{Little_Rubin_2014}. The evaluation had two components: first, we compared the performances of the three methods in estimating the model parameters $\gamma$ and $\beta$ given in Section 3.1. Second, we compared the WEE-based estimators for the average causal effect (WEE-OR, WEE-IPW, WEE-DR) with their counterparts derived from CC and MI approaches. We generated $1000$ independent datasets for each scenario, with sample sizes $n=500$ and $2000$.

\subsection{Estimators for Model Parameters}

In this subsection, we compared the performance of the proposed WEE approach and existing missing data methods in estimating the model parameters $\gamma$ and $\beta$ for two scenarios for $Y$ (binary and continuous). In both scenarios, we generated $C_1$ from a normal distribution $N(-0.5,1)$ and $A$ from a logistic regression $\operatorname{logit} \left\{\operatorname{pr}(A=1\mid c_1) \right\} = \gamma_0 + \gamma_1 c_1$. The outcome model for binary $Y$ was $\operatorname{logit} \left\{\operatorname{pr}(Y=1\mid a,c_1) \right\} = \beta_0 + \beta_1 a + \beta_2 c_1$, where the missing indicator $R_{C}$ was generated from $\operatorname{logit} \left\{\operatorname{pr}(R_{C}=1\mid c_1, y) \right\} = 0.5 - c_1 + 2y$. The outcome model for continuous $Y$ was $Y \mid a,c_1 \sim N(\beta_0 + \beta_1 a + \beta_2 c_1, 1)$, where  the missing indicator $R_{C}$ was generated from $\operatorname{logit} \left\{\operatorname{pr}(R_{C_1}=1\mid c_1, y) \right\} = -1 + c_1 + y$. The values of the parameters in the treatment propensity score model and the outcome model, $(\gamma_0, \gamma_1, \beta_0, \beta_1, \beta_2)$, were set as $(0.5, 0.5, 0.5, 1.5, -0.5)$ in both simulation scenarios. The proportion of missing data in $C_1$ was approximately $12\%$ in the binary outcome model and $48\%$ in the continuous outcome model, respectively, while $A$ and $Y$ were fully observed.  

When implementing MI, we used the predictive mean matching approach \citep{Little_Rubin_2014} for imputing the missing values. We compared the performances of WEE estimator, CC estimator, and MI estimator by calculating their bias, estimated asymptotic standard errors ($\widehat{\operatorname{Std}}$), sample standard errors (Std), and empirical coverage probabilities of the estimated $95\%$ confidence interval based on $\widehat{\operatorname{Std}}$. The simulation results are presented in Table \ref{simupara}. The proposed weighted estimating equation estimators exhibited negligible biases and the corresponding coverage probabilities of the $95\%$ confidence intervals approximated the nominal value. The sample standard errors were close to the estimated asymptotic standard errors, indicating that the large-sample estimate of variance was satisfactory. In contrast, the estimators obtained from complete-case analysis and multiple imputation displayed relatively large biases in most cases, which were not mitigated as the sample size increases. However, it is noteworthy that in the binary outcome scenario, the complete-case analysis estimator of $\beta_1$ performed well, which is consistent with the findings of \cite{Bartlett_Harel_Carpenter_2015}.

\begin{table}[htbp]
\begin{adjustwidth}{-1.5cm}{-1.5cm}
  \centering
  \begin{threeparttable}
  \caption{Simulation results for the estimators of parameters in the treatment propensity score model and the outcome model under different sample sizes}\label{simupara}
    \begin{tabular}{clrrrrrrrrrr}
    \hline
          &       & \multicolumn{2}{c}{$\gamma_0$} & \multicolumn{2}{c}{$\gamma_1$} & \multicolumn{2}{c}{$\beta_0$} & \multicolumn{2}{c}{$\beta_1$} & \multicolumn{2}{c}{$\beta_2$} \\
          \cmidrule(lr){3-4}\cmidrule(lr){5-6}\cmidrule(lr){7-8}\cmidrule(lr){9-10}\cmidrule(lr){11-12}
    \multicolumn{2}{c}{sample size} & 500   & 2000  & 500   & 2000  & 500   & 2000  & 500   & 2000  & 500   & 2000 \\
    \hline
    \\
    \multicolumn{5}{l}{Binary $Y$}\\
    \hline
    \multirow{4}[0]{*}{WEE} & Bias  & 0.001 & -0.005 & 0.000 & -0.005 & 0.021 & 0.006 & 0.042 & 0.015 & 0.048 & 0.000 \\
          & $\widehat{\operatorname{Std}}$ & 0.137 & 0.068 & 0.141 & 0.070 & 0.192 & 0.096 & 0.308 & 0.154 & 0.239 & 0.120 \\
          & Std   & 0.136 & 0.071 & 0.127 & 0.073 & 0.202 & 0.096 & 0.325 & 0.156 & 0.200 & 0.112 \\
          & CI    & 0.945 & 0.945 & 0.973 & 0.937 & 0.931 & 0.950 & 0.931 & 0.945 & 0.975 & 0.967 \\
          \hline
    \multirow{4}[0]{*}{CC} & Bias  & 0.117 & 0.116 & 0.064 & 0.067 & 0.494 & 0.497 & 0.039 & 0.017 & 0.260 & 0.258 \\
          & $\widehat{\operatorname{Std}}$ & 0.120 & 0.060 & 0.108 & 0.054 & 0.205 & 0.102 & 0.314 & 0.155 & 0.151 & 0.075 \\
          & Std   & 0.125 & 0.061 & 0.113 & 0.055 & 0.208 & 0.102 & 0.318 & 0.155 & 0.143 & 0.070 \\
          & CI    & 0.823 & 0.514 & 0.898 & 0.744 & 0.346 & 0.000 & 0.948 & 0.954 & 0.587 & 0.058 \\
          \hline
    \multirow{4}[0]{*}{MI} & Bias  & 0.070 & 0.071 & 0.050 & 0.060 & 0.162 & 0.149 & -0.151 & -0.148 & 0.285 & 0.264 \\
          & $\widehat{\operatorname{Std}}$ & 0.117 & 0.060 & 0.111 & 0.057 & 0.190 & 0.096 & 0.257 & 0.129 & 0.155 & 0.082 \\
          & Std   & 0.122 & 0.060 & 0.115 & 0.057 & 0.188 & 0.097 & 0.256 & 0.129 & 0.138 & 0.076 \\
          & CI    & 0.905 & 0.782 & 0.917 & 0.818 & 0.866 & 0.680 & 0.909 & 0.783 & 0.542 & 0.093 \\
          \hline
          \\
    \multicolumn{5}{l}{Continuous $Y$}\\
    \hline      
    \multirow{4}[0]{*}{WEE} & Bias  & 0.004 & 0.001 & -0.010 & -0.003 & 0.017 & 0.004 & -0.023 & -0.008 & -0.008 & -0.004 \\
          & $\widehat{\operatorname{Std}}$ & 0.162 & 0.081 & 0.176 & 0.088 & 0.129 & 0.064 & 0.179 & 0.090 & 0.127 & 0.064 \\
          & Std   & 0.162 & 0.078 & 0.185 & 0.088 & 0.127 & 0.065 & 0.175 & 0.091 & 0.106 & 0.058 \\
          & CI    & 0.949 & 0.962 & 0.935 & 0.956 & 0.951 & 0.949 & 0.955 & 0.943 & 0.979 & 0.965 \\
          \hline
    \multirow{4}[0]{*}{CC} & Bias  & 0.570 & 0.565 & -0.122 & -0.122 & 0.504 & 0.502 & -0.247 & -0.245 & -0.072 & -0.076 \\
          & $\widehat{\operatorname{Std}}$ & 0.154 & 0.077 & 0.151 & 0.075 & 0.113 & 0.056 & 0.130 & 0.065 & 0.060 & 0.030 \\
          & Std   & 0.157 & 0.076 & 0.153 & 0.071 & 0.111 & 0.057 & 0.130 & 0.066 & 0.060 & 0.029 \\
          & CI    & 0.038 & 0.000 & 0.874 & 0.633 & 0.007 & 0.000 & 0.525 & 0.038 & 0.782 & 0.268 \\
          \hline
    \multirow{4}[0]{*}{MI} & Bias  & -0.205 & -0.209 & -0.232 & -0.231 & 0.191 & 0.189 & -0.091 & -0.088 & -0.093 & -0.104 \\
          & $\widehat{\operatorname{Std}}$ & 0.100 & 0.049 & 0.149 & 0.073 & 0.085 & 0.042 & 0.105 & 0.052 & 0.060 & 0.029 \\
          & Std   & 0.099 & 0.049 & 0.149 & 0.071 & 0.085 & 0.041 & 0.105 & 0.052 & 0.061 & 0.030 \\
          & CI    & 0.452 & 0.016 & 0.663 & 0.113 & 0.391 & 0.006 & 0.861 & 0.612 & 0.651 & 0.060 \\
          \hline
    \end{tabular}
   \begin{tablenotes}
   \small
   \item Note: WEE, the purposed weighted estimating equation method; CC, the complete-case analysis method; MI, the multiple imputation method using predictive mean matching; $\widehat{\operatorname{Std}}$, the estimated asymptotic standard error; Std, the sample standard error; CI, the empirical coverage probability of the estimated $95\%$ confidence interval.
   \end{tablenotes}
   \end{threeparttable}
\end{adjustwidth} 
\end{table}

\subsection{Estimators for the Average Treatment Effect}

We compared the three estimators for the average causal effect (WEE-DR, WEE-IPW, WEE-OR) proposed in Section 3.2 with existing methods for handling MAR data and for causal effect estimation \citep{Bang_Robins_2005}. The existing methods we considered were (1) complete case analysis for handling missing data and outcome regression modelling for causal effect estimation (CC-OR); (2) multiple imputation for handling missing data and outcome regression modelling for causal effect estimation (MI-OR); (3) complete case analysis for handling missing data and inverse probability weighting for causal effect estimation (CC-IPW); (4) multiple imputation for handling missing data and inverse probability weighting for causal effect estimation (MI-IPW); (5) complete case analysis for handling missing data and augmented inverse probability weighting for causal effect estimation (CC-AIPW); (6) multiple imputation for handling missing data and augmented inverse probability weighting for causal effect estimation (MI-AIPW). 

We included two confounding variables: a continuous confounder $C_1$ generated from a normal distribution $N(0,1)$ and a binary confounder $C_2$ generated from a Bernoulli distribution $\operatorname{Ber}(0.5)$. 

We generated $R_{C}$ from a logistic regression $\operatorname{logit} \left\{ \operatorname{pr}(R_C=1 \mid c_1,c_2,y) \right\} = 1 - 2c_1 + c_2 + 3y$ with $R_C=1$ if $C_1$ is observed and $R_C=0$ if it is missing. It should be noted that the missing probability is directly influenced by the value of $C_1$, indicating that the missingness mechanism for $C_1$ is MNAR.

We also explored the impact of model misspecification on the performances of the proposed estimators and considered four scenarios: (a) both the outcome model and the treatment propensity score model are correctly specified (OCPC); (b) the outcome model is correctly specified and the treatment propensity score model is misspecified (OCPM); (c) the outcome model is misspecified and the treatment propensity score model is correctly specified (OMPC); (d) neither the outcome model nor the treatment propensity score model is correctly specified (OMPM). The data-generating mechanisms of the treatment $A$ and the outcome $Y$ for each of the four scenarios are detailed in Table \ref{datagen}.  In all four scenarios, we specified the treatment propensity score model as $\operatorname{logit} \left\{ \operatorname{pr}(A=1 \mid c_1,c_2) \right\} = \gamma_0 + \gamma_1 c_1 + \gamma_2 c_2$ and the outcome model as $Y \sim N(\beta_0+\beta_1 a + \beta_2 c_1 + \beta_3 c_2, \phi)$. 

\begin{table}[htbp]
  \centering
  \caption{A description of data-generating mechanisms of $A$ and $Y$}\label{datagen}
    \begin{tabular}{cl}
    \hline
    \multicolumn{1}{c}{Scenario} & \multicolumn{1}{l}{Data-generating Mechanism} \\
    \hline
    \multirow{2}[0]{*}{(a). OCPC} & $A \sim \operatorname{Ber}(\operatorname{expit}\{-0.5+c_1+c_2 \})$ \\
          & $Y \sim N(1 + 3a + c_1 - c_2 ,1 ) $ \\
    \hline
    \multirow{2}[0]{*}{(b). OCPM} & $A \sim \operatorname{Ber}(\operatorname{expit}\{-3 + 3c_1c_2 +3\exp(c_1c_2) \})$  \\
          &  $Y \sim N(1 + 3a + c_1 - c_2 ,1) $\\
    \hline
    \multirow{2}[0]{*}{(c). OMPC} & $A \sim \operatorname{Ber}(\operatorname{expit}\{-0.5+c_1+c_2 \})$ \\
          &  $Y \sim N(-1 + 3a + 0.5\exp(c_1+c_2), 1) $ \\
    \hline
    \multirow{2}[0]{*}{(d). OMPM} & $A \sim \operatorname{Ber}(\operatorname{expit}\{-3 + 3c_1c_2 +3\exp(c_1c_2) \})$  \\
          & $Y \sim N(-1 + 3a + 0.5\exp(c_1+c_2), 1) $ \\
    \hline
    \\
    \end{tabular}
\end{table}

\begin{figure}[htbp]
\begin{adjustwidth}{-0cm}{-0cm}
\centering
\includegraphics[height=8cm]{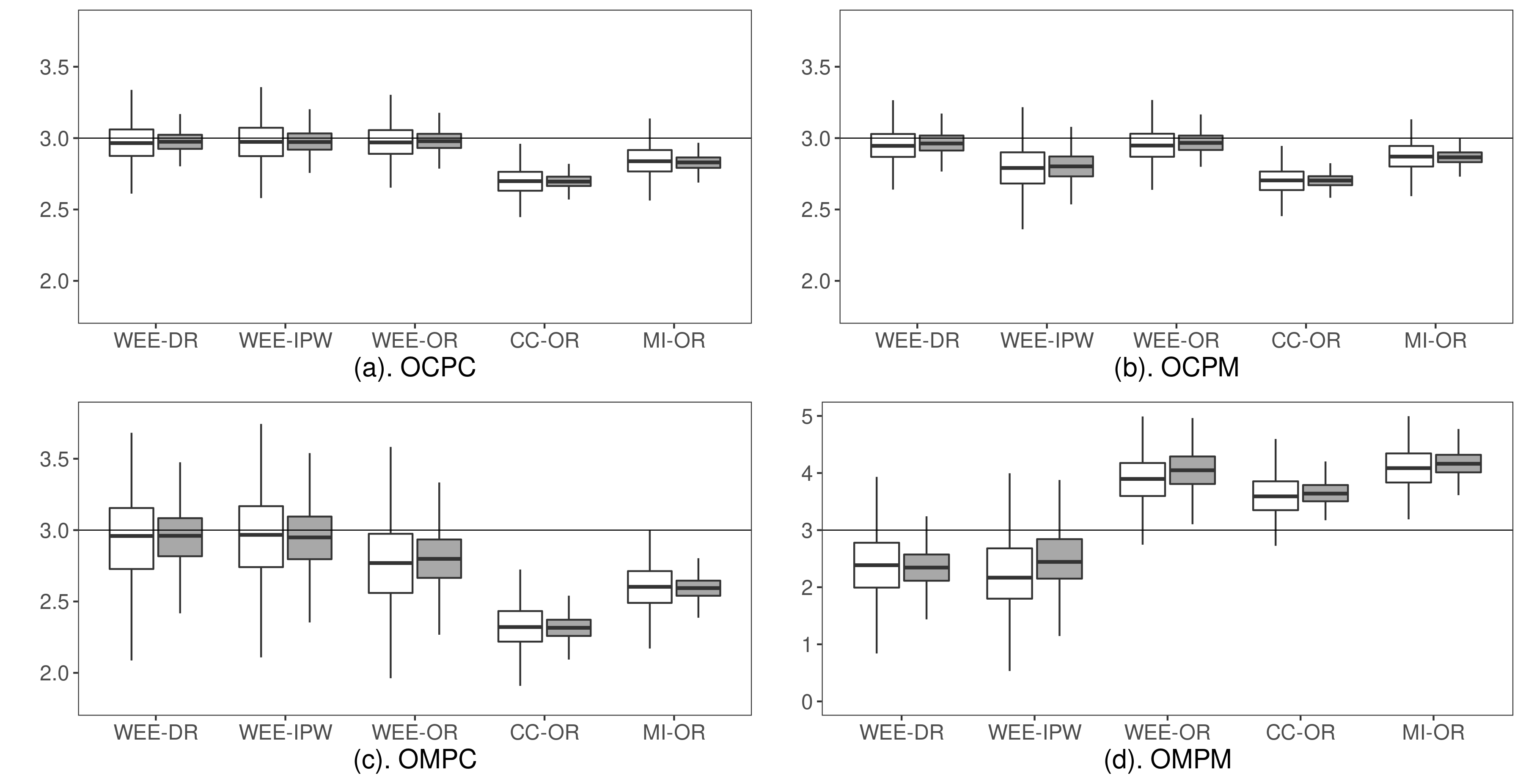}
\caption{Boxplots of average causal effect estimators. Note: In each boxplot, white boxes are for sample size 500 and gray ones for 2,000. The horizontal line marks the true value of the average causal effect.}
\label{simufig}
\end{adjustwidth}
\end{figure}

The main simulation results are presented in Figure \ref{simufig}, which displays boxplots of the WEE-DR, WEE-IPW, WEE-OR, CC-OR, and MI-OR estimators of the average causal effects. Additional results of other estimators are provided in Web Appendix K. Our simulation results show that both the CC-OR and MI-OR estimators exhibit bias across all simulated scenarios. In contrast, the proposed WEE-IPW and WEE-OR estimators are unbiased for the cases without model misspecification. However, the WEE-OR estimator is biased when the outcome model is misspecified, while the WEE-IPW estimator is biased when the treatment propensity score model is misspecified. The WEE-DR estimator demonstrates good performance in situations where either the outcome model or the treatment propensity score model, or both, are correctly specified. These findings suggest that the WEE-DR method is a robust approach to estimate the average causal effects, even when model misspecification is present.

\section{Application to Real Data}

We illustrate the proposed methods by analyzing the 2017-2018 U.S. National Health and Nutrition Examination Survey (NHANES) data to investigate the causal effect of marital status on depression \citep{Knol_et_al_2010}. Our sample consisted of 2918 individuals aged 45 years and above. Of these, 1771 were married, including living together ($A=1$), while 1147 were single, including divorced and widowed ($A=0$).  The outcome variable of interest was the Patient Health Questionnaire-9 (PHQ-9) score, which is a continuous measure of depression severity ranging from $0$ to $27$ \citep{Kroenke_Spitzer_Williams_2001}. The PHQ-9 score was calculated using a nine-item depression screening instrument from the Questionnaire Data part of the 2017-2018 NHANES dataset, where each person was asked $9$ symptom questions regarding his or her frequency of depression symptoms over the past two weeks. Following \cite{Knol_et_al_2010}, we included the monthly income-to-poverty ratio, age, and gender as potential confounders. Additionally, education level was included as a confounder in our analysis, since it may causally influence both marital status \citep{Cherlin_2010} and mental health \citep{Assari_2020}. Among the confounders included in the analysis, only the monthly income-to-poverty ratio contains missing values. It is likely that the missingness of this variable is missing not at random, as individuals with high income may be reluctant to provide their income information \citep{Davern_et_al_2005}. The missing rate of the income variable for married individuals was 19.3\%, while that for single individuals was 18.0\%. It is plausible that this missingness is independent of marital status, after adjusting for the PHQ-9 score, age, gender, education, and income information, which allows for the use of the treatment-independent missingness assumption in our analysis.

We considered a linear outcome model and a logistic regression for both the missing data model and the treatment propensity score model. We applied the proposed WEE-DR, WEE-IPW, and WEE-OR estimators as described in Section 3, and compared their performance with existing methods including CC-DR, CC-IPW, CC-OR, MI-DR, MI-IPW, and MI-OR, as described in Section 4. We used a standard bootstrap approach with 1000 resamples with replacement to obtain standard errors and confidence intervals of the estimates. Our estimated standard errors and confidence intervals were derived from the sample standard errors and the $2.5\%$ and $97.5\%$ quantiles of 1000 point estimates obtained from the resampled data, respectively.

\begin{table}[htbp]
  \centering
  \begin{threeparttable}
  \caption{Result for the real data analysis}
    \begin{tabular}{lrrr}
    \hline
          & \multicolumn{1}{c}{Est}  & \multicolumn{1}{c}{BS Std} & \multicolumn{1}{c}{BS 95CI}  \\
          \hline
    $\widehat{\alpha}_{\text{income}}$ & -1.211 & 0.179 & (-1.252 , -1.003) \\
    $\widehat{\tau}_{\text{WEE-DR}}$ & -0.441  & 0.121  & (-0.686 , -0.239) \\
    $\widehat{\tau}_{\text{WEE-IPW}}$ & -0.432  & 0.120  & (-0.678 , -0.235) \\
    $\widehat{\tau}_{\text{WEE-OR}}$ & -0.545  & 0.126  & (-0.783 , -0.320) \\
    $\widehat{\tau}_{\text{CC-AIPW}}$ & -0.885  & 0.205  & (-1.230 , -0.472) \\
    $\widehat{\tau}_{\text{CC-IPW}}$ & -0.864  & 0.203  & (-1.252 , -0.504) \\
    $\widehat{\tau}_{\text{CC-OR}}$ & -0.930  & 0.209  & (-1.319 , -0.521)\\
    $\widehat{\tau}_{\text{MI-AIPW}}$ & -0.912  & 0.190  & (-1.241 , -0.541) \\
    $\widehat{\tau}_{\text{MI-IPW}}$ & -0.896  & 0.187 & (-1.252 , -0.555) \\
    $\widehat{\tau}_{\text{MI-OR}}$ & -0.966  & 0.192  & (-1.329 , -0.555) \\
    \hline
    \end{tabular}%
    \label{RDA2}
   \begin{tablenotes}
   \small
   \item Est, point estimates; BS Std, the sample standared error of 1000 points estimates of bootstrap samples; BS 95CI, CI constracted by bootstrap quantiles; $\widehat{\alpha}_{\text{income}}$, estimate of the parameter for the monthly income-to-poverty ratio in the missing probability model;
   $\widehat{\tau}_{\text{WEE-DR}}$, $\widehat{\tau}_{\text{WEE-IPW}}$, $\widehat{\tau}_{\text{WEE-OR}}$, $\widehat{\tau}_{\text{CC-AIPW}}$, $\widehat{\tau}_{\text{CC-IPW}}$, $\widehat{\tau}_{\text{CC-OR}}$, $\widehat{\tau}_{\text{MI-AIPW}}$, $\widehat{\tau}_{\text{MI-IPW}}$, and $\widehat{\tau}_{\text{MI-OR}}$, average causal effect estimates of corresponding methods. 
   \end{tablenotes}
   \end{threeparttable}
\end{table}%

Table \ref{RDA2}  provides a summary of the estimated average causal effects obtained from these methods. The estimate for the effect of income on the missing probability, $\hat\alpha_{\text{income}}=-1.211$, was found to be significantly different from $0$, indicating that the underlying missingness mechnism is MNAR. The differences in point estimates obtained through the weighted estimating equation (WEE) method compared to complete case (CC) and multiple imputation (MI) methods highlighted the influence of missing mechnism assumptions on causal inference when dealing with missing data in confounders. Among the proposed methods, the WEE-IPW and WEE-DR estimates were similar, while the WEE-OR estimate differed slightly, suggesting a correct specification of the treatment propensity score model but not the outcome model. Based on the WEE-DR estimator, being single was found to increase the PHQ-9 score by an average of 0.441.

\section{Conclusions}

We investigate the identification and estimation of causal effects in observational studies with confounders missing not at random, under the assumption of treatment-independent missingness. In this regard, we established the identification of causal effects in some widely used parametric outcome models, treatment propensity score models, and missing probability models, and proposed weighted estimating equation-based estimators, including a doubly robust estimator, for the average causal effect. Simulation results showed that our proposed estimator remained unbiased even in the presence of mis-specified outcome models, while complete case analysis and multiple imputation methods often produced biased estimators when dealing with confounders missing not at random. Our research provides a valuable contribution to the growing body of literature on causal inference in the context of missing confounder data.


Further research may extend the proposed method to handle more complex missing data patterns beyond the current focus on single missing confounders. Another promising avenue for future research could be applying our proposed method to longitudinal studies, where missing not at random data is frequently encountered.

\end{document}